Applying of The new Integral "KAJ Transform" in Cryptography

Jinan A. Jasim

Department of Mathematics

College of Education

Mustansiriyah University

Baghdad, Iraq

Maysam Rahi Ali

College of Engineering

Al-Qadisiyah University

Al-Qadisiyah, Iraq

Emad A. Kuff

Department of Materials

College of Engineering

Al-Qadisiyah University

Al-Qadisiyah, Iraq

jinanadel78@uomustansiriyah.edu.iq    maysamrahi7@gmail.com    emad.abbas@qu.edu.iq

Corresponding author: Jinan A. Jasim; jinanadel78@uomustansiriyah.edu.iq

*Abstract*:

In this study, a new sort of transform known as (Kuffi- Abbas- Jawad transform) or KAJ- integral transformation is introduced. We introduce and explore important KAJ- transformation features and applications in cryptography. KAJ- transformation is used for encryption and inverse KAJ- transformation is used for decryption; an example is provided to illustrate the encryption and decryption of the given data.

## 1. *Introduction:*

In the past two centuries, integral transformations have been utilized efficiently to solve numerous problems in applied mathematics and engineering science. Integral transformations are well-known for their involvement in solving linear differential equations, difference equations, integral equations, and other scientific problems [1-6] due to their prominence.

A KAJ- integral transformation is employed as a security tool in coding theory for encrypting and decrypting a message in our paper's presentation. The KAJ- transformation derives from the standard Laplace integral [7]. The procedure of encrypting and decrypting methods to obtain the message is straightforward due to the mathematical simplicity of this change and its fundamental features. This transform is inspired by the Laplace, Elzaki, Sumudu, Aboodh, Kamal, and Mahgoub complex transforms. During the presentation of our methodologies, a new way of cryptography employing the KAJ- transform is offered. This approach is used to encrypt plaintext, whereas the inverse KAJ- transform is used to decrypt it. This process encrypts data by replacing each letter with its corresponding American Standard Code for Information Interchange (ASCII) value [8-15].

*Definition and Important Properties [7]:*

The new integral transformation KAJ- integral transformation (denoted by $S_m$), as a modification on the SEE integral transformation, is given as:

$$S_m^*\{h(t)\} = 1(c) = \frac{1}{c^m}\int_{t=0}^{\infty} f\left(\frac{t}{c}\right) e^{-t}\, dt,$$

Where $m$ is an integer number, $0 < L_1 \leq c \leq L_2$, where $L_1$ and $\leq L_2$ are either finite or infinite.

### *KAJ- integral transformation for some Basic functions [7]:*

1. Let $h(t) = \beta$, where $\beta$ is constant, applying KAJ- transform, we get:
   $S_m^*\{\beta\} = \frac{\beta}{c^{m+1}}$.

2. Let $h(t) = t$, $S_m^*\{t\} = \frac{1}{c^{m+1}}$.

3. Let $h(t) = t^m$, $S_m^*\{t^m\} = \frac{m!}{c^{m+m}}$, $m \in \mathbb{N}$.

4. Let $h(t) = e^{\alpha t}$, $S_m^*\{e^{\alpha t}\} = \frac{c}{c^m(c-\alpha)}$, $\alpha$ is a constant.

5. Let $h(t) = \sin(\alpha t)$, $S_m^*\{\sin(\alpha t)\} = \frac{\alpha}{c^{m-1}(c^2+\alpha^2)}$.

6. Let $h(t) = \cos(\alpha t)$, $S_m^*\{\cos(\alpha t)\} = \frac{1}{v^{m-2}(v^2+\alpha^2)}$.

7. Let $h(t) = \sinh(\alpha t)$, $S_m^*\{\sinh(\alpha t)\} = \frac{\alpha}{c^{m-1}(c^2-\alpha^2)}$.

8. Let $h(t) = \cosh(\alpha t)$, $S_m^*\{\cosh(\alpha t)\} = \frac{1}{v^{m-2}(v^2-\alpha^2)}$.

### I. *The Inverse of KAJ-Transformation [7]:*

If $S_m^*\{h(t)\} = 1(c)$ is the KAJ- transformation, then $h(t) = S_m^{-1}\{1(c)\}$ is said to be an inverse of KAJ-transform.

In this part, the inverse of KAJ-transformation of basic functions is presented:

1. $S_m^{-1}\left\{\frac{1}{c^{m+1}}\right\} = 1$.
2. $S_m^{-1}\left\{\frac{m!}{c^{m+m}}\right\} = t^m$, $m \in \mathbb{N}$.
3. $S_m^{-1}\left\{\frac{1}{c^{m-1}(c-\alpha)}\right\} = e^{\alpha t}$, $\alpha$ is a constant.
4. $S_m^{-1}\left\{\frac{\alpha}{c^{m-1}(c^2+\alpha^2)}\right\} = \sin(\alpha t)$.
5. $S_m^{-1}\left\{\frac{1}{v^{m-2}(v^2+\alpha^2)}\right\} = \cos(\alpha t)$.
6. $S_m^{-1}\left\{\frac{\alpha}{c^{m-1}(c^2-\alpha^2)}\right\} = \sinh(\alpha t)$.
7. $S_m^{-1}\left\{\frac{1}{v^{m-2}(v^2-\alpha^2)}\right\} = \cosh(\alpha t)$.

### II. *Linearity Property of KAJ-Transformation [7]:*

If $S_m^*\{h(t)\} = 1_1(c)$ and $S_m^*\{g(t)\} = 1_2(c)$, then

$$S_m^*\{\alpha h(t) \pm \beta g(t)\} = \alpha S_m^*\{h(t)\} \pm \beta S_m^*\{h(t)\},$$

Where $\alpha$ and $\beta$ are arbitrary constants.

## 2. Apply The KAJ- Transform in Cryptography:
### 2.1. Procedures for Encryption Algorithms:
- Plaintext messages should have ASCII values assigned to each letter.
- Based on the above conversion, the plaintext message is then set up as a set of numbers..
- Use the KAJ-integral transform of the polynomial $h(t) = Ft^2 e^{\beta t}$. Here, F is the plaintext message's ASCII values.
- Figure out $r_k$ so that $r_k = M_k \bmod 500$, where $1 \leq k \leq n$. Here, $M_k$ are the coefficients of $S_m*h(t)+$ from the equation above.
- Using the preceding procedure, determine a new finite series of remainders $r_1, r_2, \cdots, r_n$.
- The encrypted message is the ASCII values of $r_1, r_2, \cdots, r_n$ and the set of quotients is used as the key $C_k$, $k = 1, 2, \cdots, n$.

### 3.2. Procedures for Decryption Algorithms
- Use the specified $C_k$ to convert the given secret message back to its original text, where $k$ can be any value between 1 and $n$. And then prepare the encrypted text using the matching finite sequence of numbers, $r_1, r_2, \cdots, r_n$.
- Consider the congruence $M_k = c_k + 500 r_k$, in order to construct the initial message, where $k$ might range anywhere from 1 to $n$.
- Using the method described above, determine the values $M_1, M_2, \cdots$ and $, M_n$.
- To the $S_m*h(t)+$function, apply the inverse of the KAJ- transformation..
- Create a sequence out of the coefficients of the polynomial $h(t)$. This sequence should be finite.
- Using the ASCII values, convert the number of the finite sequence to alphabets.

After that, we have the initial message in its unencrypted form as a result of this process.

## 3. Proposed Methodology

Let us start with a plaintext having the message "ENVIRONMENT".

### A. Encryption procedure

Using the first stage of the encryption process, assign each letter of the plaintext message an appropriate value in the ASCII range

$E = 69, N = 78, V = 86, I = 73, R = 82, O = 79, N = 78, M = 77, E = 69,$

$N = 78, T = 84.$

By the second step of the encryption technique, the plaintext message is structured as a discrete sequence of ASCII values, such as the following:

$F_1 = 69, F_2 = 78, F_3 = 86, F_4 = 73, F_5 = 82, F_6 = 79, F_7 = 78, F_8 = 77,$

$F_9 = 69, F_{10} = 78, F_{11} = 84.$

The total number of terms are $n = 11$.

Now, take into consideration the typical growth of an exponential

$$e^{\beta t} = 1 + \frac{\beta t}{1!} + \frac{\beta^2 t^2}{2!} + \cdots,$$

$$t^2 e^{\beta t} = 1 + \frac{\beta t^3}{1!} + \frac{\beta^2 t^4}{2!} + \cdots$$

In accordance with the third phase of the encryption algorithm, take into consideration the polynomial.

$h(t) = Ft^2 e^{\beta t}, \quad \beta = 2.$

$$h(t) = 69 t^2 + 78 \frac{2t^3}{1!} + 86 \frac{2^2 t^4}{2!} + 73 \frac{2^3 t^5}{3!} + 82 \frac{2^4 t^6}{4!} + 79 \frac{2^5 t^7}{5!} + 78 \frac{2^6 t^8}{6!}$$
$$+ 77 \frac{2^7 t^9}{7!} + 69 \frac{2^8 t^{10}}{8!} + 78 \frac{2^9 t^{11}}{9!} + 84 \frac{2^{10} t^{12}}{10!}. \qquad (1)$$

then, apply KAJ- integral transform on both sides, we obtain:

$S_m^* h(t) + = S_m\{Ft^2 e^{\beta t}\}$

$$= \frac{69}{0!} \left(\frac{2!}{c^{m+3}}\right) + \frac{156}{1!} \left(\frac{3!}{c^{m+4}}\right) + \frac{344}{2!} \left(\frac{4!}{c^{m+5}}\right) + \frac{584}{3!} \left(\frac{5!}{c^{m+6}}\right) + \frac{1312}{4!} \left(\frac{6!}{c^{m+7}}\right)$$
$$+ \frac{2528}{5!} \left(\frac{7!}{c^{m+8}}\right) + \frac{4992}{6!} \left(\frac{8!}{c^{m+9}}\right) + \frac{9856}{7!} \left(\frac{9!}{c^{m+10}}\right) + \frac{17664}{8!} \left(\frac{10!}{c^{m+11}}\right)$$
$$+ \frac{39936}{9!} \left(\frac{11!}{c^{m+12}}\right) + \frac{86016}{10!} \left(\frac{12!}{c^{m+13}}\right). \qquad (2)$$

Now, using step four of the encryption technique, assess the remainders r k in the process of encryption in such a way that $r_k = M_k \bmod 500$, where $k$ can be any of the following values: $1, 2, 3, \ldots, n$. Where $M_k$ represents the coefficients of $S_m^* h(t)+$ given in the above polynomial of eq. (2) and modulo is taken for any given number; for example, we take modulo 500.

Step five of the encryption process is used, and a new finite series of remainders $r_1, r_2, \cdots, r_{11}$ is determined by applying the step described above in the following manner:

$r_1 = 138 \bmod 500 = 500(0) + 138 = 138.$

$r_2 = 936 \bmod 500 = 500(1) + 436 = 436.$

$r_3 = 4128 \bmod 500 = 500(8) + 128 = 128.$

$r_4 = 11680 \bmod 500 = 500(23) + 180 = 180.$

$r_5 = 39360 \bmod 500 = 500(78) + 360 = 360.$

$r_6 = 106176 \bmod 500 = 500(212) + 176 = 176.$

$r_7 = 279552 \bmod 500 = 500(559) + 52 = 52.$

$r_8 = 709632 \bmod 500 = 500(1419) + 132 = 132.$

$r_9 = 1589760 \bmod 500 = 500(3179) + 260 = 260.$

$r_{10} = 4392960 \bmod 500 = 500(8785) + 460 = 460.$

$r_{11} = 11354112 \bmod 500 = 500(22708) + 112 = 112.$

Using the sixth stage of the encryption technique, the message that has been encrypted is represented as the ASCII values of the remainders 138, 436, 128, 180, 360, 176, 52, 132, 260, 460, 112, and the set of quotients that were used to generate the key.

$c_1 = 0, c_2 = 1, c_3 = 8, c_4 = 23, c_5 = 78, c_6 = 212, c_7 = 559, c_8 = 1419, c_9 = 3179, c_{10} = 8785$ and $c_{11} = 22708.$

Because of this, the plaintext message "ENVIRONMENT" is converted into the cipher text 'èyÇ┤ Ũ░4äAffip' using the ASCII values of the remainders $r_1, r_2, \cdots,$ and $r_{11}$ (obtained by pressing Alt key of these remainders values using the keyboard of the computer).

## B. *Decryption Procedure*

Using step one of the decryption process, convert the supplied secret message to its original text by using the given key $c_k$ for $k = 1, 2, \cdots, 11$ as 0, 1, 8, 23, 78, 212, 559, 1419, 3179, 8785, 22708. This should be done by entering these values into the appropriate fields.

It has been brought to our attention that the digits 138, 436, 128, 180, 360, 176, 52, 132, 260, 460, and 112 appear in the form of a finite sequence in connection with the encrypted text.

Using the associated finite sequence of $r_1 = 138, r_2 = 436, r_3 = 128, r_4 = 180, r_5 = 360, r_6 = 176, r_7 = 52, r_8 = 132, r_9 = 260, r_{10} = 460$ and $r_{11} = 112.$

Determine the values of $M_1, M_2, \cdots,$ and so on up to $M_{11}$ by applying step two of the decryption process and utilizing the step above. Then we get

$M_1 = 138, M_2 = 936, M_3 = 4128, M_4 = 11680, M_5 = 39360, M_6 = 106176, M_7 = 279552, M_8 = 709632, M_9 = 1589760, M_{10} = 4392960$ and $M_{11} = 11354112.$

By using eq. (2), we obtain:

$$S_m\{h(t)\} = 138\left(\frac{1}{c^{m+3}}\right) + 936\left(\frac{1}{c^{m+4}}\right) + 4128\left(\frac{1}{c^{m+5}}\right) + 11680\left(\frac{1}{c^{m+6}}\right)$$
$$+ 39360\left(\frac{1}{c^{m+7}}\right) + 106176\left(\frac{1}{c^{m+8}}\right) + 279552\left(\frac{1}{c^{m+9}}\right)$$
$$+ 709632\left(\frac{1}{c^{m+10}}\right) + 1589760\left(\frac{1}{c^{m+11}}\right) + 4392960\left(\frac{1}{c^{m+12}}\right)$$
$$+ 11354112\left(\frac{1}{c^{m+13}}\right).$$

After carrying out the fourth phase of the decryption procedure, which involves carrying out the inverse of the KAJ-integral transform, we obtain:

$$h(t) = (69)t^2 + (78)\frac{2t^3}{1!} + (86)\frac{2^2 t^4}{2!} + (73)\frac{2^3 t^5}{3!} + (82)\frac{2^4 t^6}{4!} + (79)\frac{2^5 t^7}{5!}$$
$$+ (78)\frac{2^6 t^8}{6!} + (77)\frac{2^7 t^9}{7!} + (69)\frac{2^8 t^{10}}{8!} + (78)\frac{2^9 t^{11}}{9!}$$
$$+ (84)\frac{2^{10} t^{12}}{10!}.$$

Using the fifth step of the decryption process, arrange the coefficients of the polynomial h(t) in a finite sequence. For example, you could do something like this: 69,78,86,73,82,79,78,77,69,78,84.

By applying the ASCII values to the integers in the finite sequence and converting them to alphabets in step six of the decryption technique, we are able to recover the original text message "ENVIRONMENT."

## *Conclusion:*

Encryption and decryption of the supplied message are both accomplished through the use of the KAJ-integral transform of exponential functions with ASCII values. In our work with cryptography, we made use of some of the useful aspects of this transformation. The algorithmic portion is not difficult at all. Due to the fact that this operation involves a high value of modulus, the plain text message in safety form is permitted (500). Because of this, the process of encrypting and decrypting data is made much more secure.

## *References:*

[1]. Davies, B., Integral transform and their Applications. Second Edition, Springer Science & Business Media, LLc, (1985).
[2]. Dharshini Priya S., Muthu Amirtha & Senthil Kumar P., Application of N-transform in Cryptography. International Journal of Scientific Research and Reviews, 8(1), 2143-2147, (2019).
[3]. Hemant, K. Undegaonkar, Security in Communication by using Laplace Transforms and Cryptography. International Journal of Scientific & Technology Research, Volume 8, Issue 12, 3207-3209, (2019).
[4]. Kenneth, H. Rosen, Discrete Mathematics and Its Applications. McGraw Hill, (2012).
[5]. Debnath, L., & Bhatta, D., Integral Transforms and Their Applications. Second Edition, Chapman & Hall/CRC, (2006).
[6]. Mohand, M., & Mahgoub, A., The new integral transform "Mohand Transform". Advances in Theoretical and Applied Mathematics, 12(2), 113-120, (2017).

[7]. E.S. Abbas, E.A. Kuffi, A.A. Jawad, New Integral Kuffi-Abbas-Jawad KAJ Transform and its application on Ordinary Differential Equations, Journal of Interdisciplinary Mathematics, 25(5), 1427-1433, (2022), https://doi.org/10.1080/09720502.2022.2046339.

[8]. Burton D. M., Elementary Number Theory. Tata McGraw Hill, New Delhi, (2002).

[9]. Erwin Kreyszig, Advanced Engineering Mathematics. John Wiley and Sons Inc, (1999).

[10]. Johannes A. Buchmann, Introduction to Cryptography. Springer, (2004).

[11]. Semwal, P., & Sharma, M. K., Comparative Study of Different Cryptographic Algorithms for Data Security in Cloud Computing. International Journal on Emerging Technologies (Special Issue NCETST) 8(1), 746-750, (2017).

[12]. Amiruddin, A., Ratna, A. A. P., & Sari, R. F., Systematicreview of internet of things security. International Journal of Communication Networks and Information Security, 11(2), 248-255, (2019).

[13]. Yadav, S. K., & Kumar, K. (2011). On Certain attacks and Privacy-Protecting Coupon System. International Journal of Theoretical and Applied Sciences, 3(2), 79-87.

[14]. Tarig M. Elzaki. (2011). The New Integral Transform "Elzaki Transform". Global Journal of Pure and Applied Mathematics, 7(1), 57-64.

[15]. Srinivas V., Jayanthi C.H, Application of the New Integral "J-transform" in Cryptography, International Journal on Emerging Technologies, 11(2), p.p.: 678-682, 2020.